\documentclass[prl,preprint,amsmath,amssymb,aps]{revtex4}
\usepackage{graphicx}
\usepackage[justification=centering]{caption}
\usepackage{amsmath}
\usepackage{amssymb}
\usepackage{lipsum}
\usepackage{dsfont}
\usepackage{multirow}
\usepackage{afterpage}
\usepackage{amsfonts}

\begin{document}

\title{Affinity-based measurement-induced nonlocality and its robustness against noise}
\author{R. Muthuganesan}
\author{V. K. Chandrasekar}

\affiliation{Centre for Nonlinear Science \& Engineering, School of Electrical \& Electronics Engineering, SASTRA Deemed University, Thanjavur - 613 401, Tamil Nadu, India}

\bigskip

\begin{abstract}
Measurement-induced nonlocality (MIN), a quantum correlation measure for the bipartite system, is an indicator of global effects due to locally invariant von Neumann projective measurements. It is well-- known fact that the correlation measures based on Hilbert-Schmidt norm  are not credible measure in capturing nonlocal attributes of a quantum state. In this article, to remedy the local ancilla problem of  Hilbert-Schmidt norm based MIN, we propose a new form of MIN based on affinity. This quantity satisfies all criteria of a bonafide measure of quantum correlation measure. For an arbitrary pure state, it is shown that affinity based MIN equals to other form of geometric versions of correlation measure. We obtain an upper bound of this measure for $m \times n$ dimensional arbitrary mixed state.  We obtain a closed formula of the proposed version of MIN for $2 \times n$ dimensional (qubit--qudit) mixed state. We apply these results on two--qubit mixed states such Werner, isotropic and Bell diagonal state. To illustrate the robustness of affinity--based measure against noise, we study the dynamics of MIN under generalized amplitude damping channel.
\end{abstract}

\maketitle

\section{\bf Introduction}~

Quantum correlation, a fundamental aspect of quantum mechanics and significantly makes the departure from the classical regime. It is a useful physical resource for various quantum information processing such as teleportation, super dense coding, communication, and quantum algorithm \cite{Nielsen2010}. Quantification of correlation between its local constituents of a system is a formidable task in the framework of quantum information theory. In this regard, entanglement is believed as a valid resource for quantum advantageous in the early $\text{19}^{th}$ century \cite{Bell, ES}. In the light of seminal work of Werner \cite{Werner1989} and the presence of non-zero quantum correlation namely, discord (beyond entanglement) \cite{Ollivier2001}, it is believed that entanglement is alone not responsible for the advantageous over the classical algorithm. Recently, various measures have been proposed to capture quantumness, which goes beyond entanglement such as measurement-induced disturbance (MID) \cite{Luo2008}, geometric discord (GD) \cite{Dakic2010, Luo2010pra}, measurement-induced nonlocality (MIN) \cite{Luo2011} and uncertainty-induced nonlocality (UIN) \cite{UIN}.

The local disturbance is an important tool to probe nonlocal aspects of a quantum state. In particular, locally invariant von Neumann projective measurements can induce global or nonlocal effects. Using  local projectors of marginal states various measures have been studied extensively using local eigenprojectors \cite{Luo2008,Luo2010pra,Luo2011,UIN}. Distance between the quantum state in state space is useful in quantification of quantum correlation. Geometric measures are easy to compute and experimentally realizable \cite{Dakic2010}.  In this context, several distance measures have been introduced, few such measures are trace distance, Hilbert-Schmidt norm, Jensen-Shannon divergence, fidelity induced metric and Hellinger distance \cite {Bengtsson,Jozsa2015}. Recently, measurement based nonclassical correlations (a family of discord-like measures) have been characterized using various distance measures \cite{Spehner}.  

MIN, characterize the nonlocality of a quantum state in the perspective of locally invariant projective measurements, is  more general than the Bell nonlocality. It is also a bona fide measure of quantum correlation for the bipartite state. MIN has also been investigated based on relative entropy \cite{Xi2012}, von Neumann entropy, skew information \cite{Li2016}, trace distance \cite{Spehner, Hu2015} and fidelity \cite{Muthu1}.  In this article, we introduce a new variant of MIN based on affinity. It is shown that this quantity is a remedy for the local ancilla problem of Hilbert-Schmidt norm based MIN. This measure satisfies all criteria of a bona fide measure of quantum correlation of bipartite system. Affinity--based MIN is a valid resource for  quantum communication, cryptography and dense coding.  Further, we evaluate the affinity--based MIN analytically for arbitrary pure state. We obtain an upper bound for $m \times n$ dimensional mixed state  and a closed formula of the proposed  MIN for the $2 \times n$ dimensional mixed state. We study the quantumness of well-known family of two-qubit $m \times m$ dimensional mixed states. Finally, we study the dynamical behavior of affinity based measure under a noisy quantum channel.

\section{Measurement-induced nonlocality}
Measurement-induced nonlocality, captures nonlocal or global effect of a quantum state due  to von Neumann projective measurements, is originally defined as maximal square of Hilbert-Schmidt norm of the difference of pre- and post- measurement states. Mathematically, it is defined as \cite{Luo2011}
\begin{equation}
 N(\rho ) =~^{\text{max}}_{\Pi ^{A}}\| \rho - \Pi ^{A}(\rho )\| ^{2},  \label{HS-MIN}
\end{equation}
Here the maximum is taken over the von Neumann projective measurements on subsystem $A$, $\Pi^{A}(\rho) = \sum _{k} (\Pi ^{A}_{k} \otimes   \mathds{1} ^{B}) \rho (\Pi ^{A}_{k} \otimes    \mathds{1}^{B} )$, and $\Pi ^{A}= \{\Pi ^{A}_{k}\}= \{|k\rangle \langle k|\}$ being the projective measurements on the subsystem $A$, which do not change the marginal state $\rho^{A}$ locally i.e., $\Pi ^{A}(\rho^{A})=\rho ^{A}$. The dual of this quantity is geometric discord (GD) of the given state $\rho$ formulated as \cite{Luo2010pra}
\begin{equation}
 D(\rho ) =~^{\text{min}}_{\Pi ^{A}}\| \rho - \Pi ^{A}(\rho )\| ^{2}. 
\end{equation}
For nondegenerate $\rho^{A}$ the optimization is not required. Hence, both the MIN and GD are equal. Due to the computability and easy accessibility of experimentation, the researchers paid much attention to the both GD and MIN. However, both the quantities are having an unwanted property of  quantum correlation measure \cite{Piani2012}, it may change rather arbitrarily through some trivial and uncorrelated action on the unmeasured party $B$. 

Consider a simple mapping $\Gamma ^\sigma: X\rightarrow X \otimes\sigma$, i.e., the map adding a noisy ancillary state to party $B$. Under such an operation 
\begin{align}
\| X \| \rightarrow \| \Gamma ^\sigma X \|=\| X \| \sqrt{\text{Tr}\sigma^2}.
\end{align}
Since the Hilbert-Schmidt norm is multiplicative on tensor products. After the addition of local ancilla $\rho^C$, MIN of the resultant state is computed as 
\begin{align}
 N(\rho^{A:BC} ) = N(\rho^{AB}) \text{Tr}(\rho ^{C})^2  \nonumber
\end{align}
implying that MIN differs arbitrarily due to local ancilla $C$  as long as $\rho^{C}$ is mixed. Since the optimization over the projective measurements on $A$ is unaffected by the presence of uncorrelated ancilla state on $B$ (unmeasured party). Thus, adding or removing local ancilla -- a local and reversible operation. Due to this operation, a factor namely, purity of the ancillary state is added to the original MIN.

A natural way to resolve this local ancilla problem is to redefine the MIN (\ref{HS-MIN}) as \cite{Piani2012}
\begin{align}
\tilde{N}(\rho) =~^{\text{max}}_{\Lambda_{B}}~~ N(\rho^{AB})_{\Lambda_B}, 
\end{align}
where the maximum is over the channel $\Lambda_B$.
On the other way, Luo and Fu remedied this local ancilla problem by replacing density matrix by its square root i.e., mathematically remedied MIN is defined as \cite{Chang2013}
\begin{equation}
 N(\rho ) =~^{\text{max}}_{\Pi ^{A}}\| \sqrt{\rho} - \Pi ^{A}(\sqrt {\rho} )\|^{2} .
\end{equation}
Further, the contractive distance measures such as trace distance, Hellinger distance and fidelity induced metric are also useful in resolve this local ancilla problem. In what follows, we define a new variant of MIN using affinity induced metric.

\section{$\alpha$- Affinity and MIN}
Metric in state space, quantify closeness or similarity of two states in the state space, they play a central role for the classification of states in information theory. They are also associated with geometric measures. Further, they are useful to quantify how precisely a quantum channel can transmit information. Affinity, like fidelity \cite{Jozsa2015} characterizes the  closeness of two quantum states. Here, we define a metric in state space and introduce a new quantum correlation measure using affinity between pre- and post- measurement states.  Classically, affinity is defined as \cite{Luo2004}
\begin{align}
  \mathcal{A}(g,h)=\sum_x(\sqrt{g(x)}\sqrt{h(x)}),
\end{align}
where $g$ and $h$ are discrete  probability distributions. This definition is alike the Bhattacharyya coefficient between two probability distributions (discrete or continuous) in classical probability theory \cite{Bhattacharyya}. Classical affinity quantifies the closeness of two probability distributions. Extending the same notion in the quantum regime, one can replace probability distribution by density matrix and the summation by trace operator. Hence, the affinity of two quantum state is defined as 
\begin{align}
  \mathcal{A}(\rho,\sigma)=\text{Tr}\left(\sqrt{\rho}\sqrt{\sigma}\right).
\end{align}
The affinity is much similar to fidelity \cite{Jozsa2015}, describes how close two quantum states are. It is also possess all the  properties of fidelity. This quantity is more useful in quantum detection and estimation theory.  The notion of affinity has been extended to $\alpha-$affinity $(0<  \alpha < 1)$, which is defined as 
\begin{align}
  \mathcal{A}_\alpha(\rho,\sigma)=\text{Tr}(\rho^ {\alpha} \sigma^{1-\alpha})
\end{align}
with $ \alpha \in(0,1)$. The $\alpha-$affinity satisfies the following properties:
\begin{enumerate}
\item[(i)] $\alpha-$affinity is bounded i.e., $0 \leq \mathcal{A}_\alpha(\rho,\sigma)\leq 1$ and $\mathcal{A}_\alpha(\rho,\sigma)=1$ if and only if $\rho=\sigma$ for all values of $\alpha$.
\item[(ii)] Monotonic $(\mathcal{A}_\alpha(\Phi(\rho),\Phi(\sigma))\geq  \mathcal{A}_\alpha(\rho,\sigma))$ under completely positive and trace preserving (CPTP) map.
\item[(iii)] Joint concavity $\mathcal{A}_\alpha(\sum_ip_i\rho_i,\sum_ip_i\sigma_i)\geq \sum_ip_i\mathcal{A}_\alpha(\rho_i,\sigma_i)$.
\end{enumerate}

In general, affinity itself is not a metric. Due to monotonicity and concavity property of affinity \cite{Luo2004}, one can define  any monotonically decreasing function of affinity  as a metric in state space. One such affinity-based metric is
\begin{align}
  d_{\mathcal{A}_\alpha}(\rho,\sigma)=\sqrt{1-\mathcal{A}_\alpha(\rho,\sigma)}. \label{metric}
\end{align}
Further, it is easy to show that the above--defined metric satisfies all the axioms of a valid distance measure in state space. To define affinity based measurement--induced nonlocality, we set $\alpha=1/2$. For simplicity, here onwards we drop subscript $\alpha$ and we denote the  $1/2-$affinity $(\mathcal{A}_{1/2})$ as $\mathcal{A}$.
Defining MIN in terms of affinity using above-mentioned metric as 
\begin{align}
N_{\mathcal{A}}(\rho)=~^\text{{max}}_{ \Pi^{A}} ~  d^2_{\mathcal{A}}(\rho,\Pi^A(\rho))=1-~^\text{{min}}_{ \Pi^{A}} ~\text{Tr}\left(\sqrt{\rho}\sqrt{\Pi^A(\rho)}\right),
\end{align}
where the maximum/minimum is taken over von Neumann projective measurements. In principle, one can generalize this definition for the multipartite scenario. Using the identity $\Pi^A f(\rho)\Pi^A=f(\Pi^A \rho \Pi^A)$ \cite{{Girolami2012}}, one can rewrite the definition of MIN using affinity as 
\begin{align}
N_{\mathcal{A}}(\rho)=~1-^\text{{min}}_{ \Pi^{A}} \text{Tr}\left[ \sqrt{\rho} \Pi^A(\sqrt{\rho})\right].  \label{identity}
\end{align}
It is worthful to mention that this quantity satisfies all necessary axioms of quantum correlation measure. Here, we demonstrate some interesting properties of affinity--based MIN: 
\begin{enumerate}
\item[(i)]  $N_\mathcal{A}(\rho)$ is non-negative i.e., $N_\mathcal{A}(\rho)\geq 0$. 

\item[(ii)] $N_\mathcal{A}(\rho)=0$ for any product state $\rho=\rho_{A}\otimes  \rho _{B}$ and the classical-quantum state in the form $\rho =\sum _{k}p_{k}|k\rangle \langle k| \otimes   \rho_{k}  $ with nondegenerate marginal state $\rho^{A}=\sum_{k}p_{k}|k\rangle \langle k|$.

\item[(iii)] $N_\mathcal{A}(\rho)$ is locally unitary  invariant i.e., $N_\mathcal{A}\left((U\otimes   V)\rho  (U\otimes   V)^\dagger\right)=N_\mathcal{A}(\rho)$ for any local unitary operators $U$ and $V$. 
\item[(iv)] For any $m \times n$ dimensional pure maximally entangled state with $m\leq n$, $N_\mathcal{A}(\rho) $ has the maximal value of $\frac{m-1}{m}$. 
\item[(v)] For nondegenerate $\rho^A$, the affinity--based MIN $N_{\mathcal{A}}(\rho)= d^2_{\mathcal{A}}(\rho,\Pi^A(\rho))$.
\item[(vi)] $N_\mathcal{A}(\rho)$ is invariant under the addition of any local ancilla to the unmeasured party. 
\end{enumerate}
To prove this invariant property, we show that affinity is unaltered by the addition of uncorrelated ancilla to unmeasured party $B$. First of all, we recall the multiplicative property of affinity and is given as  
\begin{align}
\mathcal{A}(\rho_1 \otimes \rho_2,\sigma_1 \otimes \sigma_2)=\mathcal{A}(\rho_1 \otimes\sigma_1)\cdot \mathcal{A}(\rho_2 \otimes\sigma_2). \label{multiaffinity}
\end{align}
After the addition of local ancilla to unmeasured party $B$, the affinity between the pre-- and post--measurement states is
\begin{align}
\mathcal{A}\left(\rho^{A:BC},\Pi ^{A}(\rho^{A:BC})\right) = \mathcal{A}\left(\rho^{AB}\otimes  \rho ^{C},\Pi ^{A}(\rho^{AB})\otimes  \rho ^{C}\right). \nonumber
\end{align}
Using multiplicativity property of affinity Eq.(\ref{multiaffinity}), we have 
\begin{align}
 \mathcal{A}\left(\rho^{A:BC},\Pi ^{A}(\rho^{A:BC})\right) =& \mathcal{A}\left(\rho^{AB}, \Pi ^{A} (\rho ^{AB})\right)\cdot \mathcal{A}(\rho^{C},\rho^{C}), \nonumber  
\end{align}
\begin{align}
=\mathcal{A}\left(\rho^{AB}, \Pi ^{A} (\rho ^{AB})\right),
\end{align}
which completes the proof of  property  (vi). Hence, $N_{\mathcal{A}}(\rho)$ is a good measure of nonlocal correlation or quantumness in a given system.
\section{\bf MIN for Pure State}
{\bf Theorem:} \textit{For any pure bipartite state with Schmidt decomposition $| \Psi \rangle =\sum_{i}\sqrt{s_{i}}| \alpha _{i} \rangle \otimes | \beta _{i}\rangle $,}
\begin{equation}
 N_{\mathcal{A}}(| \Psi \rangle\langle \Psi| )=1- \sum_{k} s_{k}^{2}.
\end{equation}
{\bf Proof: }
Noting that 
\begin{equation}
\sqrt{\rho}= | \Psi \rangle \langle \Psi| = \sum_{ij}\sqrt{s_{i}s_{j}}| \alpha_{i} \rangle \langle \alpha_{j}| \otimes  | \beta_{i} \rangle \langle \beta_{j}|. \nonumber  
\end{equation}
The local projective measurements on the subsystem $A$ can be expressed as $\Pi ^{A}=\{\Pi ^{A}_{k}\otimes \mathds{1}\} = \{| \alpha_{k}\rangle \langle \alpha_{k}|\otimes| \mathds{1}\} $, which do not alter the marginal state $\rho^{A}$. The marginal state $\rho^A=\Pi^A(\rho^A)=\sum_k\Pi_k^A\rho^A\Pi_k^A$ is written as 
\begin{equation}
 \rho^{A}=\sum_{k} U| \alpha_{k}\rangle \langle \alpha_{k}| U^{\dagger}\rho^{A} U| \alpha_{k}\rangle \langle \alpha_{k}| U^{\dagger}.  \nonumber
\end{equation}
The spectral decomposition of the marginal state $\rho^{A}$  in the orthonormal bases $\{U| \alpha_{k}\rangle\} $ can be written as
 \begin{equation}
 \rho^{A}= \sum_{k} \langle \alpha_{k}| U^{\dagger}\rho^{A} U| \alpha_{k}\rangle U| \alpha_{k}\rangle \langle \alpha_{k}| U^{\dagger},    \nonumber
\end{equation}
where $\langle \alpha_{k}| U^{\dagger}\rho^{A} U| \alpha_{k}\rangle=s_k$ are the eigenvalues of state $\rho^A$. After a straight forward calculation and simplification, we show that 
\begin{align}
\sum_k\text{Tr}[ \sqrt{\rho} (\Pi_k^{A}\otimes \mathds{1})\sqrt{\rho}(\Pi_k^{A}\otimes \mathds{1})]=\sum_k(\langle \alpha_{k}| U^{\dagger}\rho^{A} U| \alpha_{k}\rangle)^2=\sum_k s_k^2.  \label{resultT1}
\end{align}
Substituting Eq.(\ref{identity}) in Eq.(\ref{resultT1}), we obtain the affinity based geometric discord for the pure state as,
\begin{equation}
 N_{\mathcal{A}}(| \Psi \rangle\langle \Psi| )=1- \sum_{k} s_{k}^{2}
\end{equation}
and hence theorem is proved. It is worthful to mention that for pure state, the affinity--based measure is equal to earlier quantities such as skew information, Hilbert-Schmidt norm based MINs, geometric measure of entanglement and remedied version of MIN. For pure $m \times n-$ dimensional entangled state with $m\leq n$, the quantity $\sum_{k}s^{2}_{k}$ is bounded by $1/m$. Then,
\begin{equation}
N_{\mathcal{A}}\left(| \Psi \rangle \langle \Psi|\right)\leq \frac{m-1}{m}
\end{equation}
and the equality holds for maximally entangled state.

\section{\bf MIN for Mixed state}
Let $\{X_{i}:i=0,1,2,\cdots,m^{2}-1\} \in \mathcal{L}(\mathcal{H}^A)$ be a set of orthonormal operators for the state space $\mathcal{H}^A$ with operator inner product $\langle X_{i}| X_{j}\rangle = \text{Tr}(X_{i}^{\dagger}X_{j})$. Similarly, one can define $\{Y_{j}:j=0,1,2,\cdots,n^{2}-1\}  \in \mathcal{L}(\mathcal{H}^B)$ for the state space $\mathcal{H}^B$. The operators $X_{i}$ and $Y_{j}$ are satisfying the conditions $\text{Tr}(X_{k}^{\dagger }X_{l})=\text{Tr}(Y_{k}^{\dagger}Y_{l})=\delta _{kl}$. With this, one can construct a set of orthonormal operators $\{X_{i} \otimes Y_{j} \}\in \mathcal{L} (\mathcal{H}^{A}\otimes \mathcal{H}^{B}) $ for the composite system. Consequently, an arbitrary $m\times n$ dimensional state of a bipartite composite system can be written as
\begin{equation}
\sqrt{\rho}= \sum_{i,j}\gamma _{ij}X_{i}\otimes  Y_{j}, \label{c}
\end{equation}
where $\gamma _{ij} =\text{Tr} (\sqrt{\rho} ~X_{i}\otimes  Y_{j})$ are real elements of  matrix $\Gamma$. After a straight forward algebra, one can compute the affinity between pre- and post- measurement state as 
\begin{align}
\mathcal{A}(\rho,\Pi^A(\rho))=\text{Tr}(R\Gamma\Gamma^tR^t).
\end{align}
The affinity--based MIN is 
\begin{align}
N_\mathcal{A}(\rho)=1-~^\text{{min}}_{R}~\text{Tr}(R\Gamma\Gamma^tR^t),
\end{align}
where the matrix  $R=(r_{ki})=\text{Tr}(| k\rangle \langle k|X_{i})$. Now we have,
\begin{align}
\sum_{i=0}^{m^{2}-1}r_{ki}r_{k^{'}i}=\text{Tr}\left(| k\rangle \langle k| k^{'}\rangle \langle k^{'}| \right)=\delta _{kk^{'}} \nonumber
\end{align}
with $r_{k0}=1/\sqrt{m}$. For $k=k^{'}$
\begin{align}
\sum_{i=1}^{m^{2}-1}r_{ki}^{2}= \frac{m-1}{m} \label{equalk} 
\end{align}
and for $k\neq k^{'}$
\begin{align}
\sum_{i=1}^{m^{2}-1}r_{ki}r_{k^{'}i}= -\frac{1}{m}.  \label{notequalk}
\end{align}
From Eqs. (\ref{equalk}) and (\ref{notequalk}) we can write the matrix $RR^{t}$~as
\begin{align}\nonumber
RR^{t}=\frac{1}{m}
\begin{pmatrix}
m-1 &  -1 & \cdots & -1 \\
-1 &  m-1 & \cdots & -1 \\
\vdots & \vdots & \ddots & \vdots \\
-1 &  -1 & \cdots & m-1 
\end{pmatrix},
\end{align}
which is a square matrix of order $m$ with eigenvalues $0$ and $1$ (with multiplicity of $m-1$). For this symmetric matrix, we have the similarity transformation $RR^{t}=U D U^{t}$ with real unitary operator $U$ and diagonal matrix $D$. Now constructing $m \times m^{2}$ matrix $B$ as
\begin{align}\nonumber
B=U^{t}R=
\begin{pmatrix}
R_0 \\
0 
\end{pmatrix},
\end{align}
where $R_0$ is a $(m-1) \times m^{2}$ matrix, such that $R_0R_0^{t}=\mathds{1}_{m-1}$ and we have,
\begin{align}
^\text{{min}}_{R}~\text{Tr}~(R\Gamma \Gamma ^{t}R^{t})~=~ ^\text{{min}}_{R_0}~\text{Tr}~(R_0\Gamma \Gamma ^{t}R_0^{t}). \nonumber
\end{align}
Then
\begin{align}
N_{\mathcal{A}}(\rho)=~1-~^\text{{min}}_{R_0}~\text{Tr}(R_0\Gamma \Gamma ^{t} R_0^{t}).
\end{align}
Since
\begin{align}
^\text{{min}}_{R_0:R_0R_0^{t}=~\mathds{1}_{m-1} }\text{Tr}(R_0\Gamma \Gamma ^{t} R_0^{t}) = \sum_{i=1}^{m-1}\mu _{i}, \nonumber
\end{align}
where $\mu _{i}$ are eigenvalues of the matrix $\Gamma \Gamma ^{t}$ listed in increasing order, we have the following tight upper bound for affinity based MIN as 
\begin{align}
N_{\mathcal{A}}(\rho)\leq 1-\sum_{i=m}^{m^2-1}\mu _{i}.
\end{align}

Next, we compute the closed formula of MIN for $2\times n$ dimensional state. The projective measurement operator $\Pi_k^A$ can be expressed in Bloch sphere representation, 
\begin{equation}
\Pi_k^A=\frac{1}{2}\left( \mathds{1}_2+\vec{r}\cdot \vec{\sigma}\right), \label{projector}
\end{equation}
where $\mathds{1}_2$ is $2 \times 2$ unit matrix $\vec{r}\cdot \vec{\sigma}=\sum_{i=1}^3 r_k^i \sigma_i$ with $\sum_{i=1}^3 \left(r_k^i\right)^2=1$. Substituting Eq.(\ref{projector}) in Eq. (\ref{identity}), one can arrive
\begin{equation}
N_{\mathcal{A}}(\rho)=1-^\text{{min}}_{r_k^i}~\sum_{i,j}r_k^iT_{ij}r_k^j=1-\lambda_{\text{min}},
\end{equation}
where $\lambda_{\text{min}}$ is the minimal eigenvalues matrix $T$ with matrix elements $T_{ij}=\text{Tr}\left[\sqrt{\rho}(\sigma_i\otimes \mathds{1})\sqrt{\rho}(\sigma_j\otimes \mathds{1})\right]$. Therefore $N_{\mathcal{A}}(\rho)$ can be analytically solved for any qubit-qudit states, which is different from other nonclassical correlations such as quantum discord which has no analytical formula even for two-qubit states. Interestingly, for this qubit-qudit case, the quantum correlation happens to be the local quantum uncertainty as the minimum skew information achievable on a single local measurement.
\section{Examples}
We compute the proposed quantity for well-known two--qubit Bell diagonal state, Werner state and isotropic state and compare with original version of Hilbert-Schmidt distance based MIN \cite{Luo2010}.

\textit{Bell diagonal state}: The Bloch representation of the state can be expressed as 
\begin{equation}
\rho^{BD}=\frac{1}{4}\left[\mathds{1}\otimes\mathds{1}+\sum^3_{i=1}c_i(\sigma^i \otimes \sigma^i)\right],
\end{equation}
where the vector $\vec{c}=(c_1,c_2,c_3) $ is a three dimensional vector composed of correlation coefficients such that $-1\leq c_i=\langle \sigma^i \otimes\sigma^i \rangle \leq 1$ completely specify the quantum state and $\lambda_{a,b}$,  here $a, b\in \{ 0,1\}$ denotes the eigenvalues of Bell diagonal state which are given by
\begin{equation}
\lambda_{a,b}=\frac{1}{4}\left[1+(-1)^a c_1-(-1)^{a+b}c_2+(-1)^b c_3\right]. \nonumber
\end{equation}
and $|\beta_{ab}\rangle =\frac{1}{\sqrt{2}}[ |0,b\rangle +(-1)^a|1,1+b\rangle ] $ are the Bell states. If $\rho^{BD}$ describes a valid physical state, then $0\leq \lambda_{a,b}\leq 1$ and $\sum_{a,b}\lambda_{a,b}=1$. Under this constraint, the vector $\vec{c}=(c_1, c_2, c_3) $ must be restricted to the tetrahedron whose vertices  are $(1, 1, -1)$, $(-1, -1, -1)$, $(1, -1, 1)$ and $(-1, 1, 1)$ \cite{Sarandy2013}. The vertices are easily identified as Bell states (EPR pairs), for which the correlation measures are maximum. Further, the measures are vanishing for the correlation vector $\vec{c}=(0, 0, 0) $, at which the state $\rho^{BD}=\mathds{1}/4$ is a maximally mixed state.

 To compute affinity--based measure, we first calculate the square root of the state $\rho^{BD}$ is 
\begin{align}
\sqrt{\rho^{BD}}=\frac{1}{4}\left[h \mathds{1}\otimes\mathds{1}+\sum^3_{i=1}d_i(\sigma^i \otimes \sigma^i)\right], \nonumber
\end{align}
where $h=\text{Tr}(\sqrt{\rho^{BD}})=\sum_{ab}\sqrt{\lambda_{ab}}$ and 
\begin{eqnarray}
d_1&=&\sqrt{\lambda_{00}}-\sqrt{\lambda_{01}}+\sqrt{\lambda_{10}}-\sqrt{\lambda_{11}}, \\ \nonumber
d_2&=&-\sqrt{\lambda_{00}}+\sqrt{\lambda_{01}}+\sqrt{\lambda_{10}}-\sqrt{\lambda_{11}}, \\ \nonumber
d_3&=&\sqrt{\lambda_{00}}+\sqrt{\lambda_{01}}-\sqrt{\lambda_{10}}-\sqrt{\lambda_{11}}. \\ \nonumber
\end{eqnarray}
From the definition of MIN, we compute the affinity--based MIN for Bell diagonal state is 
\begin{align}
N_{\mathcal{A}}(\rho^{BD})=1-\frac{1}{4}(h^2+\text{min}_j\{ d_j^2\}). \label{affresult}
\end{align}
In particular, if $c_1=c_2=c_3=-p$, then the Bell diagonal state reduced to two--qubit Werner state,
\begin{align}
\rho^{BD}=\frac{1-p}{4}\mathds{1}+p|\Phi\rangle \langle \Phi|,  ~~~~~~~~~~ p\in[-1/3, 1], \nonumber
\end{align}
with $|\Phi\rangle \langle \Phi|$. The MINs of the Werner state are computed as 
\begin{align}
N_{\mathcal{A}}(\rho^{BD})=\frac{1}{4}[ 1+p-\sqrt{(1-p)(1+3p)}],~~~~~~~~~~N(\rho^{BD}) =\frac{p^2}{2}.
\end{align}
\begin{figure*}[!ht]
\centering\includegraphics[width=0.6\linewidth]{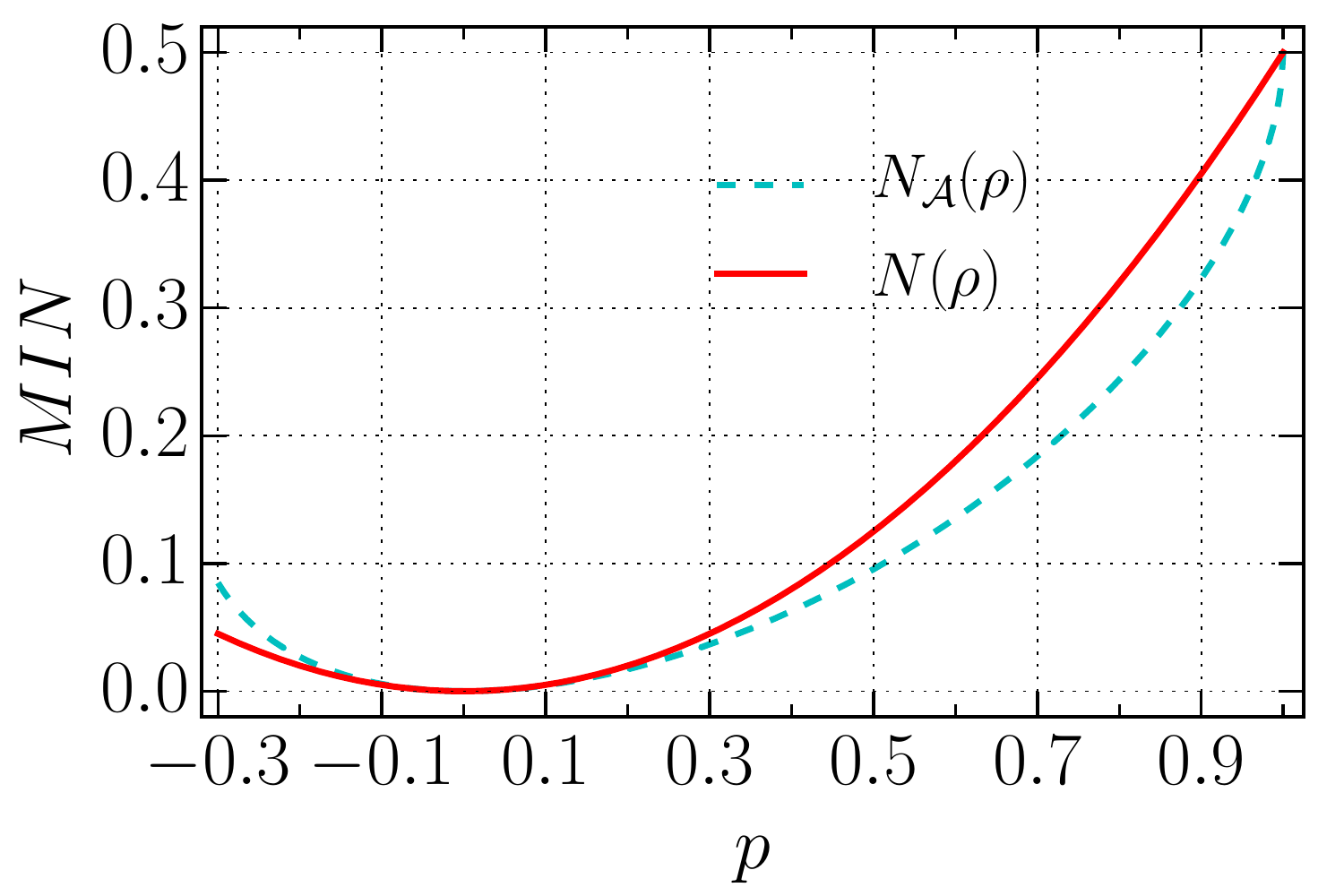}
\caption{(color online) Hilbert-Schmidt norm and affinity--based MIN for $2\times 2$ isotropic states.}
\label{fig1}
\end{figure*}
\textit{Werner state}: We consider $m\times m-$ dimensional Werner state, which is defined as \cite{Werner1989}
\begin{align}
\omega=\frac{m-\text{x}}{m^3-m}\mathds{1}+\frac{m\text{x}-1}{m^3-m}F,  ~~~~~\text{with}~~~~ \text{x}\in[-1,1],
\end{align}
with $F=\sum_{kl}|kl\rangle \langle kl|$. Affinity--based MIN is computed as 
\begin{align}
N_{\mathcal{A}}(\omega)=\frac{1}{2} \left(\frac{m-\text{x}}{m+1} -\sqrt{\frac{m-1}{m+1}(1-\text{x}^2)}\right) \nonumber
\end{align}
and Hilbert-Schmidt norm based MIN \cite{Luo2010}
\begin{align}
N(\omega)=\frac{(m\text{x}-1)^2}{m(m-1)(m+1)^2}. \nonumber
\end{align}
It is observed that $N_{\mathcal{A}}(\omega)=N(\omega)=0$, if and only if $\text{x}=1/m$. In the asymptotic limit $m\rightarrow \infty $,
\begin{align}
\lim_{m \to \infty}N_{\mathcal{A}}(\omega)=\frac{1}{2}(1-\sqrt{1-x^2}),~~~~~~~~~~\lim_{m \to \infty} N(\omega) =0. \nonumber
\end{align}
The above equation suggests that affinity--based measure is more robust in higher dimension than  the Hilbert-Schmidt norm based MIN.

\textit{Isotropic state}: $m\times m-$ dimensional isotropic state is defined as \cite{Horodecki1999}
\begin{align}
\rho=\frac{1-\text{x}}{m^2-1}\mathds{1}+\frac{m^2\text{x}-1}{m^2-1}|\Psi^+\rangle \langle \Psi^+|  ~~~~~\text{with}~~~~\text{x}\in[0,1],
\end{align}
where $|\Psi^+\rangle=\frac{1}{\sqrt{m}}\sum_i |ii\rangle $. The affinity--based MIN is computed as 
\begin{align}
N_{\mathcal{A}}(\rho)=\frac{1}{m} \left(\sqrt{(m-1)\text{x}}-\sqrt{\frac{1-\text{x}}{m+1}}\right)^2, \nonumber
\end{align}
and the original MIN is \cite{Luo2010}
\begin{align}
N(\rho)=\frac{(m^2\text{x}-1)^2}{m(m-1)(m+1)^2}. \nonumber
\end{align}
We see that $N_{\mathcal{A}}(\rho)=N(\rho)=0$ if $\text{x}=1/m^2$. In the asymptotic limit,
\begin{align}
\lim_{m \to \infty}N_{\mathcal{A}}(\rho)=\text{x},~~~~~~~~~~\lim_{m \to \infty} N(\rho) =\text{x}^2. \nonumber
\end{align}
\begin{figure*}[!ht]
\centering\includegraphics[width=0.6\linewidth]{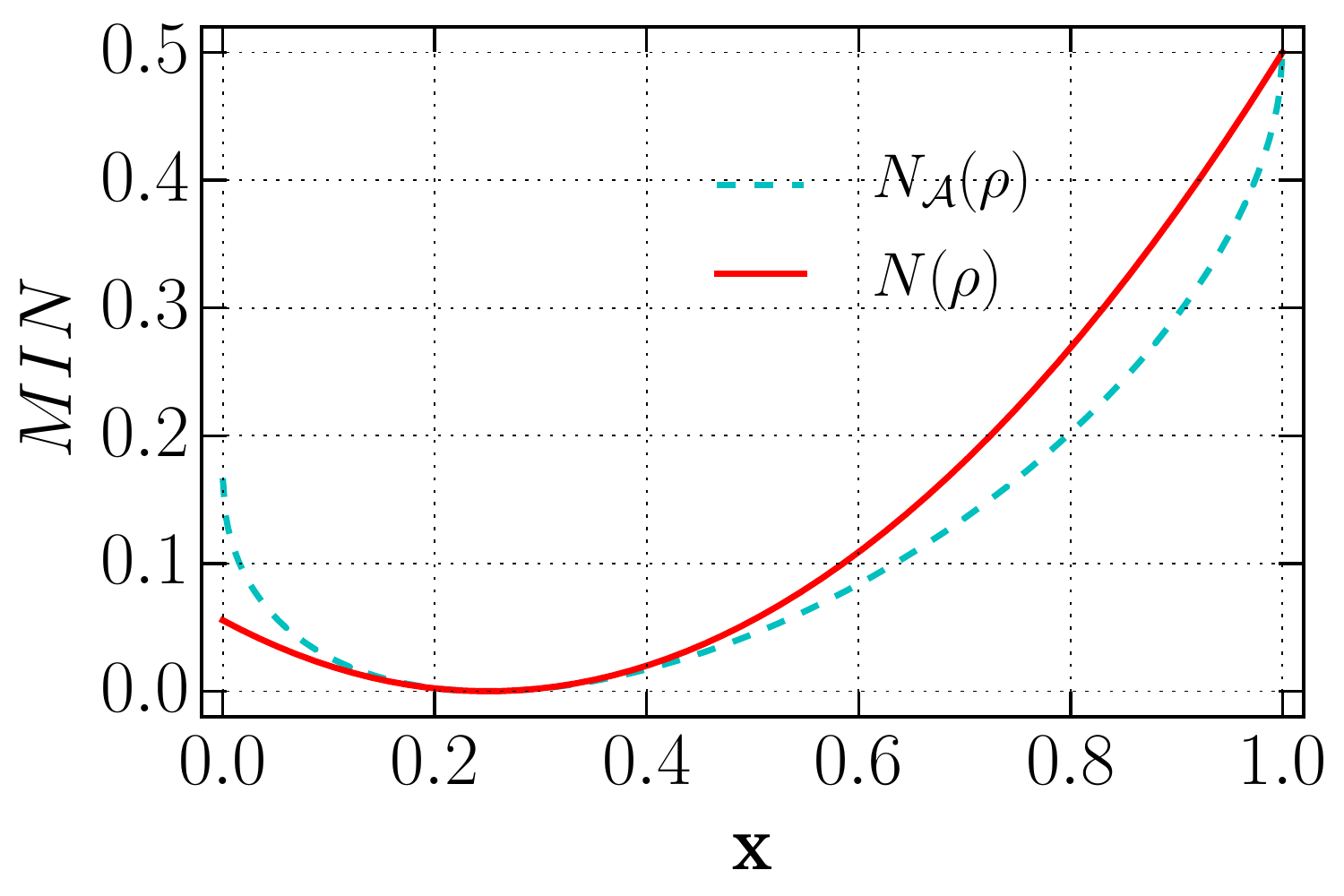}
\caption{(color online) Hilbert-Schmidt norm and affinity--based MIN for $2\times 2$ isotropic states}
\label{fig2}
\end{figure*}
Our result for $m=2$ is plotted in Fig. \ref{fig2}, which shows the consistency of affinity--based correlation with the earlier one.
\section{Dynamics}
In general, any quantum system always coupled with the environment which influences the time evolution or dynamical changes of the system. The interaction of the quantum system with the environment can be conveniently investigated using operator-sum representation. In this formalism, the evolution of quantum state is described by positive and trace preserving operation
\begin{equation}
\mathcal{E}(\rho)=\sum_{i,j}(E_i\otimes E_j) \rho (E_i\otimes E_j)^{\dagger} \label{Qoperation}
\end{equation}
where $\{ E_k\} $ is a set of Kraus operators associated with the decohering process of a single qubit and satisfy the completeness property $\sum_kE_kE^{\dagger}_k=\mathds{1}$. For an appropriate set of $E_k$, the operator-sum representation is equivalent to the dynamics of the master equation approach. 

Here, we consider the Bell diagonal state as an initial for investigation. We shall note that Bell diagonal state preserves its structure after the intervention of environment as described above. Time evolved state is then given by
\begin{equation}
\mathcal{E}(\rho^{BD})=\frac{1}{4}
\begin{pmatrix}
1+c'_3 & 0 & 0 & c'_1-c'_2 \\
0 & 1-c'_3 & c'_1+c'_2 & 0  \\
0 & c'_1+c'_2 & 1-c'_3 &0  \\
c'_1-c'_2 & 0 & 0 & 1+c'_3
\end{pmatrix}
\end{equation}
with the correlation vector $\vec{c'}=(c'_1, c'_2, c'_3)$. Here the primed components are time dependent and unprimed are initial conditions.  

\begin{figure*}[!ht]
\centering\includegraphics[width=0.48\linewidth]{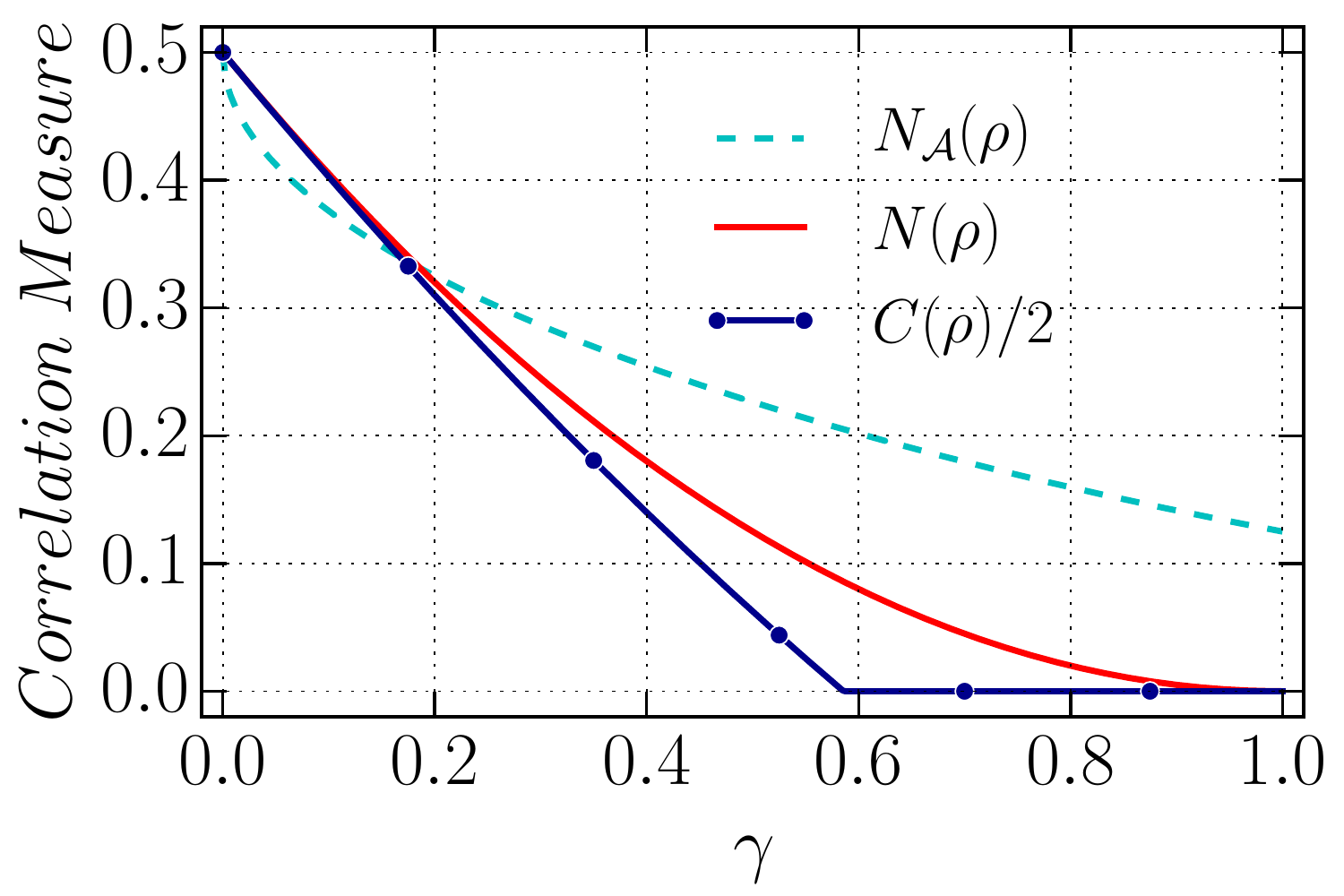}
\centering\includegraphics[width=0.48\linewidth]{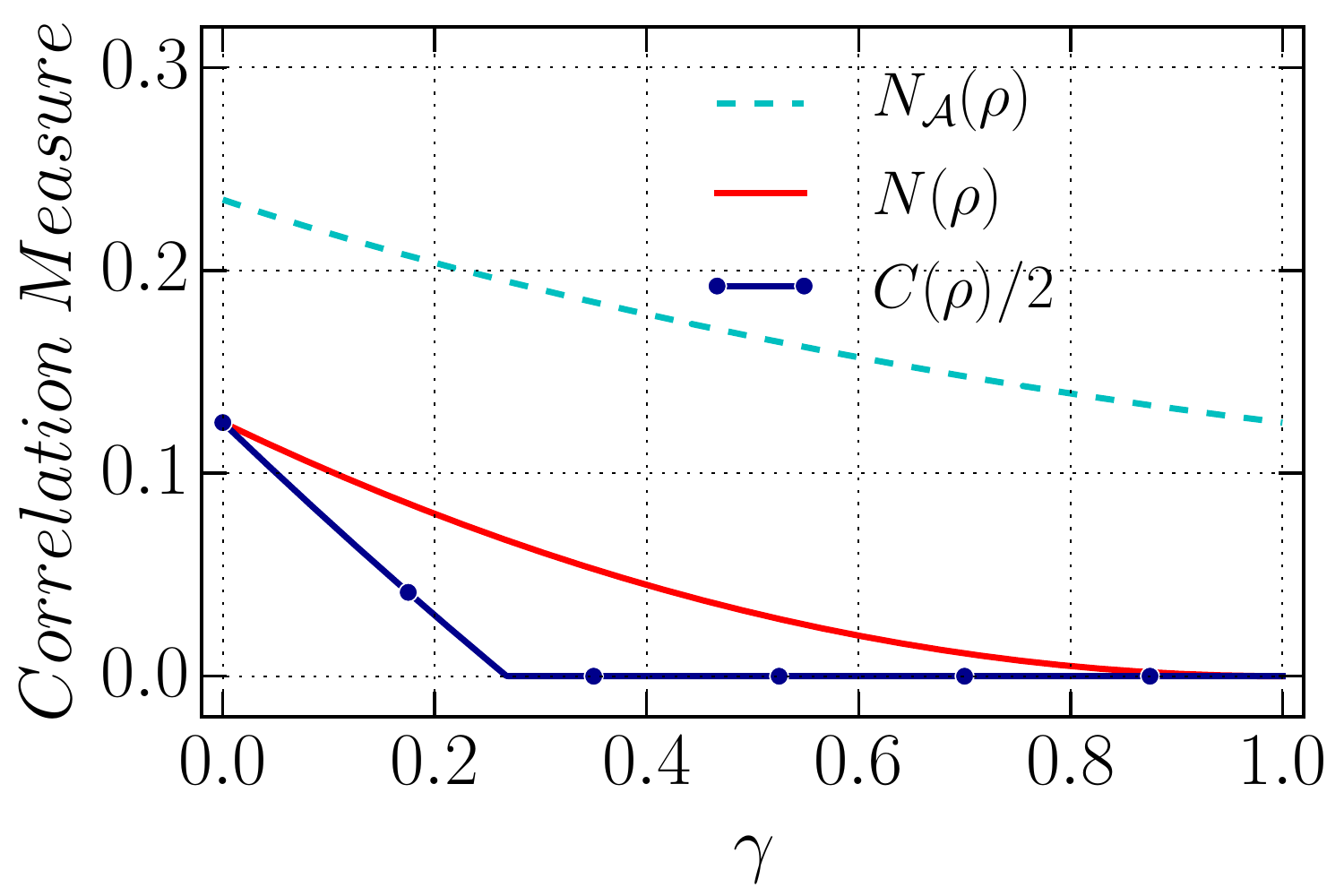}
\caption{(color online) Dynamics of quantum correlations under GAD channel for the initial  pure maximally entangled state with  $\vec{c}=(1,1,-1)$ (left) and mixed partially entangled state $\vec{c}=(0.5,0.5,-0.5)$ (right).}
\label{fig3}
\end{figure*}

\textit{Generalized Amplitude Damping:}
Here we consider the generalized amplitude damping (GAD), which models the decay of an atom from an excited state due to coupling with the environment at a finite temperature such as thermal bath. Such a process is described by the following Kraus operators \cite{Nielsen2010}
\begin{eqnarray}
E_{0}&=&\sqrt{p}
\begin{pmatrix}
1 & 0\\
0 & \sqrt{1-\gamma }
\end{pmatrix},
~E_{1}=\sqrt{p}
\begin{pmatrix}
0 & \sqrt{\gamma }\\
0 & 0
\end{pmatrix}, \nonumber \\
E_{2}&=&\sqrt{1-p}
\begin{pmatrix}
\sqrt{1-\gamma} & 0\\
0 &  1
\end{pmatrix},
~E_{3}=\sqrt{1-p}
\begin{pmatrix}
0 & 0\\
 \sqrt{\gamma } & 0
\end{pmatrix}, \nonumber
\end{eqnarray} 
where $\gamma =1-\mathrm{e}^{-\gamma' t}$, $\gamma' $ is decay rate and $p$ defines the final probability distribution of stationary (equilibrium) state. Here, we fix $p=1/2$ and the components of the evolved state under this channel are given by
\begin{equation}
  c'_1=(1-\gamma)c_1,~~~ c'_2=(1-\gamma)c_2, ~~~~c'_3=(1-\gamma)^2 c_3.
\end{equation}
Using Eq. (\ref{affresult}), one can calculate the affinity--based MIN of time evolved state by replacing primed variables. For the initial state with correlation vector $\vec{c}=(0, 0, 0) $, the coordinates of time evolved state is  $\vec{c'}=(0, 0, 0) $ i.e., $\mathcal{E}(\rho^{BD})=\rho^{BD}$ and all the correlation measures are zero. In order to understand the dynamical behavior of MIN and entanglement (measured by concurrence \cite{Hill1997}), we consider pure and maximally entangled state with the correlation vector $\vec{c}=(1, 1, -1)$. Here, we observe that all the measures are decreasing with increase of $\gamma$ and influence of quantum noise reduces the entanglement to zero in finite time. It is observed that GAD noise cause zero entanglement between the qubits for $\gamma\geq \gamma_0\simeq 0.58$ as shown in Fig.\ref{fig3}. This phenomenon is known as the sudden death of entanglement \cite{Eberly2009}. On the other hand, the companion quantities (MINs) are non-zero where the region of entanglement is zero.
Next, we consider the partially entangled state with vector $\vec{c}=(0.5, 0.5, -0.5) $ as an initial state for our analysis. One can observe similar effect as in earlier case and entanglement of the evolved state is suffered by sudden death. In this case also the nonzero MINs in the region of zero concurrence show the existence of quantum correlation without entanglement. In the light of above, we conclude that MINs are more robust than the entanglement measure against GAD channel.

\section{Conclusions}
In this article, we have proposed a new variant of measurement induced nonlocality (MIN) using affinity  metric as a measure of quantum correlation for bipartite state. It is shown that, in addition to capturing global nonlocal effect of a state due to von Neumann projective measurements, this quantity can be remedying local ancilla problem of Hilbert--Schmidt norm based MIN. We have presented a closed formula of affinity--based MIN for an arbitrary $2\times n$ dimensional mixed state, with an upper bound for $m\times n$ dimensional system. Further, we have computed the proposed version of MIN for $m \times m$ dimensional Werner and isotropic states. Finally, we have studied the dynamics of MINs and compared with entanglement. It is shown that MINs are more robust against decoherence than the entanglement.

\bigskip
\noindent{\bf Acknowledgments}\, This work is supported by the CSIR EMR Grant No.03(1444)/18/EMR-II.


\begin{thebibliography}{99}
\bibitem{Nielsen2010}
M. Nielsen, I. Chuang, Quantum Computation and Quantum Information, Cambridge (2010).

\bibitem{Bell}
J. S. Bell, Physics {\bf 1} (1964) 195.

\bibitem{ES}
E. Schrodinger, Proc. Cambridge Philos. Soc. {\bf 32} (1936) 446.

\bibitem{Werner1989}
R.F. Werner, Phys. Rev. A {\bf 40} (1989) 4277.

\bibitem{Ollivier2001}
H. Ollivier, W.H. Zurek, Phys. Rev. Lett. {\bf 88} (2001) 017901.

\bibitem{Luo2008}
S. Luo, Phys. Rev. A. {\bf 77} (2008) 022301.

\bibitem{Dakic2010}
Daki\'c, B. Vedral, V. Brukner, Phys. Rev. Lett. {\bf 105} (2010) 190502.

\bibitem{Luo2010pra}
S. Luo, S. Fu, Phys. Rev. A {\bf 82} (2010) 034302.

\bibitem{Luo2011}
S. Luo, S. Fu, Phys. Rev. Lett. {\bf 106} (2011) 120401.

\bibitem{UIN}
S.-Xiong, W. J. Zhang, C. -S Yu,  H-S. Song, Phys. Lett. A {\bf 378} (2014) 344.

\bibitem{Bengtsson}
I. Bengtsson, K. Zyczkowski.: Geometry of Quantum states:  An introduction to Quantum entanglement (Cambridge University Press, Cambridge, England, 2006).

\bibitem{Jozsa2015}
R. Jozsa,  J. Mod. opt. {\bf 41} (1994) 2315.

\bibitem{Spehner}
D. Spehner, F. Illuminati, M. Orszag, W. Roga.: Lectures on General Quantum Correlations and their Applications. Springer (2017).

\bibitem{Xi2012}
Z. Xi, X. Wang, Y. Li, Phys. Rev. A {\bf 85} (2012) 042325.

\bibitem{Li2016}
L. Li, Q.W. Wang, S.Q. Shen, M. Li, Euro. Phys. Lett. {\bf 114} (2016) 10007.

\bibitem{Hu2015}
M.-L. Hu,  H. Fan, New J. Phys. {\bf 17} (2015) 033004.

\bibitem{Muthu1}
R. Muthuganesan, R. Sankaranarayanan, Phys. Lett. A {\bf 381} (2017) 3028.

\bibitem{Piani2012}
M. Piani, Phys. Rev. A {\bf 86} (2012) 034101.

\bibitem{Chang2013}
L. Chang, S. Luo, Phys. Rev. A {\bf 87} (2013) 062303. 

\bibitem{Luo2004} 
S. Luo, Q. Zhang,  Phys. Rev. A {\bf 69} (2004) 032106.

\bibitem{Bhattacharyya}
A. Bhattacharyya, Bulletin of the Calcutta Mathematical Society {\bf 35} (1943) 99.

\bibitem{Girolami2012}
D. Girolami, G. Adesso, Phys. Rev. Lett. {\bf 108} (2012) 150403. 

\bibitem{Luo2010}
S. Luo, S. Fu, Euro. phys. Lett. {\bf 92} (2010) 20004.

\bibitem{Horodecki1999}
M. Horodecki, P. Horodecki, Phys. Rev. A {\bf 59} (1999) 4206.

\bibitem{Sarandy2013}
J. D. Montealegre, F. M. Paula, A. Saguia,  M. S. Sarandy,  Phys. Rev. A, {\bf 87}, 042115 (2013).

\bibitem{Hill1997}
 S. Hill, W. K. Wootters, Phys. Rev. Lett. {\bf 78}, 5022-5025 (1997).

\bibitem{Eberly2009}
Ting Yu, J. H. Eberly,  Science, {\bf 293}, 5914 (2009).


\end{thebibliography}
\end{document}